# Crystallization seeds favour crystallization only during initial growth

E. Allahyarov[1,2], K. Sandomirski[3], S.U. Egelhaaf[3] & H. Löwen[1]

Crystallization represents the prime example of a disorder–order transition. In realistic situations, however, container walls and impurities are frequently present and hence crystallization is heterogeneously seeded. Rarely the seeds are perfectly compatible with the thermodynamically favoured crystal structure and thus induce elastic distortions, which impede further crystal growth. Here we use a colloidal model system, which not only allows us to quantitatively control the induced distortions but also to visualize and follow heterogeneous crystallization with single-particle resolution. We determine the sequence of intermediate structures by confocal microscopy and computer simulations, and develop a theoretical model that describes our findings. The crystallite first grows on the seed but then, on reaching a critical size, detaches from the seed. The detached and relaxed crystallite continues to grow, except close to the seed, which now prevents crystallization. Hence, crystallization seeds facilitate crystallization only during initial growth and then act as impurities.

[1] Institute for Theoretical Physics II: Soft Matter, Heinrich Heine University, Universitätsstrasse 1, 40225 Düsseldorf, Germany. [2] Theoretical Department, Joint Institute for High Temperatures of the Russian Academy of Sciences, Izhorskaya Street, 13 Boulevard 2, Moscow 125412, Russia. [3] Condensed Matter Physics Laboratory, Heinrich Heine University, Universitätsstrasse 1, 40225 Düsseldorf, Germany. Correspondence and requests for materials should be addressed to S.U.E. (email: Stefan.Egelhaaf@uni-duesseldorf.de).

In addition to its fundamental importance[1–9], crystallization is relevant for many applications. They range from material science, including metals[10], small molecules[11], colloids[12] and photonics[13], to complex plasmas[14], meteorology[15], medicine[16,17] and biotechnology, where the crystallization of proteins represents an important issue[17–19]. Furthermore, everyday-life phenomena, for example, cloud condensation[15], the icing on airplane wings[20] and the weathering of rocks[21], involve crystallization. In practical and industrial situations, container walls and impurities are usually present and hence heterogeneous nucleation dominates[12,15,19–27]. Heterogeneous nucleation is thus often unavoidable and sometimes also desired to enhance crystallization. It can be induced by individual seed particles[15,22], particle assemblies[23,24], structured walls[12,19,25,26] or flat walls[27]. However, typically 'real' seeds do not perfectly match the thermodynamically favoured crystal structure, for example, owing to a different unit cell structure or size. For an only modest mismatch, the system initially follows the usual heterogeneous crystallization scenario, although with a slower crystallization speed[8,22,28–30], similar to the slower crystallization observed in the presence of polydispersity[31]. On the other hand, in the case of a very strong mismatch, the seeds no longer favour crystallization, but act as impurities and suppress crystallization[32]. Thus, if in a practical situation crystallization is observed, it is very likely that a mismatch is present, which is however small enough to still allow for crystallization. Therefore, some mismatch is expected to affect many crystallization processes in nature and in industry, but also represent a scientifically interesting issue due to the intricate interplay between different driving forces. Nevertheless, this process is hardly understood on the single-particle level.

Combining experiments and simulations, we follow the sequence of intermediate structures during heterogeneous crystallization in a colloidal model system. The crystallite starts to grow heterogeneously on the seed. Owing to seed-induced distortions, elastic stress accumulates in the crystallite. To relax the elastic stress, the crystallite detaches from the seed on reaching a critical size. The detached and relaxed crystallite continues to grow in bulk. However, a thin fluid layer remains close to the seed, which now prevents crystallization and hence acts as an impurity. These findings are consistently described by a newly developed theoretical model, which we present as well. The observed scenario is expected to be independent of the specific seed and particles. It thus is anticipated to occur whenever there is no perfect match between seed and crystallite, which frequently is the case.

## Results

**Crystallization is followed on the single-particle level.** We investigated a colloidal model system in which the mismatch between seed-induced and thermodynamically favoured crystal structures can be easily tuned: colloidal hard-sphere-like particles[33] with diameter $\sigma_1$ to which a small number of large spheres with diameter $\sigma_2$ was added. In the experiments, fluorescently labelled polymethylmethacrylate (PMMA) spheres and large glass beads were used. Using confocal microscopy, they can be visualized and followed on the single-particle level (Fig. 1). From stacks of confocal slices, the coordinates of the individual particles can be determined[34] (Fig. 2) and crystalline particles (shown in red) distinguished from fluid particles (shown in blue) using local bond-orientational order parameters[35].

Confocal microscopy experiments and Brownian dynamics simulations consistently indicate that in this situation, crystallization proceeds in several stages (Fig. 2). Heterogeneous crystallization is initiated on the surface of the large sphere, which hence acts as a seed. The curvature of the seed surface results in curved crystal planes. As will be shown below, this leads to elastic distortions that accumulate during the crystallite's growth. To release the elastic stress, the crystallite detaches from the seed surface once it reaches a critical size. As a consequence, the region between the seed and the detached crystallite melts to become a fluid. The fluid subsequently refreezes when the relaxed crystallite continues to grow in the bulk. However, bulk crystallization ceases before reaching the large sphere, which prevents complete crystallization and hence now acts as impurity. Qualitatively identical behaviour is found for a broad range of seed sizes, for the present conditions up to $\sigma_2 = 31\sigma_1$, whereas no detachment is observed for larger seed sizes, that is, smaller seed curvatures.

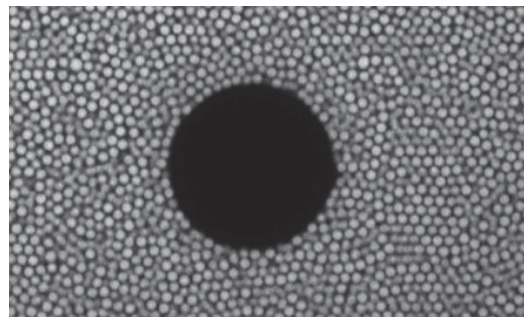

**Figure 1 | Imaging on the single-particle level.** Confocal microscopy image showing a single slice through a sample containing fluorescently labelled PMMA spheres, with diameter $\sigma_1 = 1.83\,\mu m$ and volume fraction $\Phi = 0.53$, crystallizing in the presence of a spherical glass bead of diameter $\sigma_2 = 15\sigma_1 = 28\,\mu m$ at time $t \approx 260\tau_B$ after crystallization has been started, where $\tau_B = \sigma_1^2/D_0 = 33\,s$ is the Brownian time and $D_0$ the short-time infinite-dilution diffusion coefficient.

The observed scenario indicates an intricate balance between heterogeneous crystallization on the seed and in bulk with the large sphere acting as crystallization-enhancing seed as well as crystallization-impeding impurity. The large sphere changes its role when the crystallite detaches from the seed, which hence represents the crucial turning point between heterogeneous, that is, seed induced, and bulk crystallization.

**Distinct heterogeneous and bulk growth regimes.** The different stages can be quantitatively distinguished and characterized based on the evolution of the fraction of crystalline particles $N_c(l,t)/N$ in the layers surrounding the seed, where $l$ is the layer number. The thickness of one layer is taken to be $0.91\sigma_1$, which corresponds to the separation of the crystal planes.

First, the time dependence of $N_c(l,t)/N$ in the layers close to the seed is considered (Fig. 3a,b). Initially, $N_c(l,t)/N$ increases, reflecting the growing crystallite (Fig. 2a→b). This starts in the layers very close to the seed and then extends to the layers further from the seed, leading to the crossing of $N_c(l,t)/N$ for different $l$ in the heterogeneous growth regime. The maxima of $N_c(l,t)/N$ for the first four layers indicate the end of the heterogeneous growth of the crystallite. The subsequent decrease of $N_c(l,t)/N$, especially in the layers next to the seed, reflects the detachment of the crystallite and the formation of a fluid between the seed and crystallite (Fig. 2b→c). Then $N_c(l,t)/N$ increases again, with the increase starting and being most pronounced in the layers further from the seed. This indicates the growth of the detached crystallite and (partial) refreezing of the fluid layer (Fig. 2c→d). The different stages are more pronounced in the simulations and their durations in the experiments and simulations are different. This is attributed to the hydrodynamic interactions present in the experimental system but could also be due to the different polydispersities[31], possible differences in the effective volume fractions[36], particle–particle interactions[33], and arrangements and mobilities of the particles next to the seed. Nevertheless, both, experiments and simulations, reveal the same scenario: increase, decrease and again increase of $N_c(l,t)/N$ with well-developed maxima and minima, which separate the different stages of heterogeneous crystal growth, melting and refreezing.

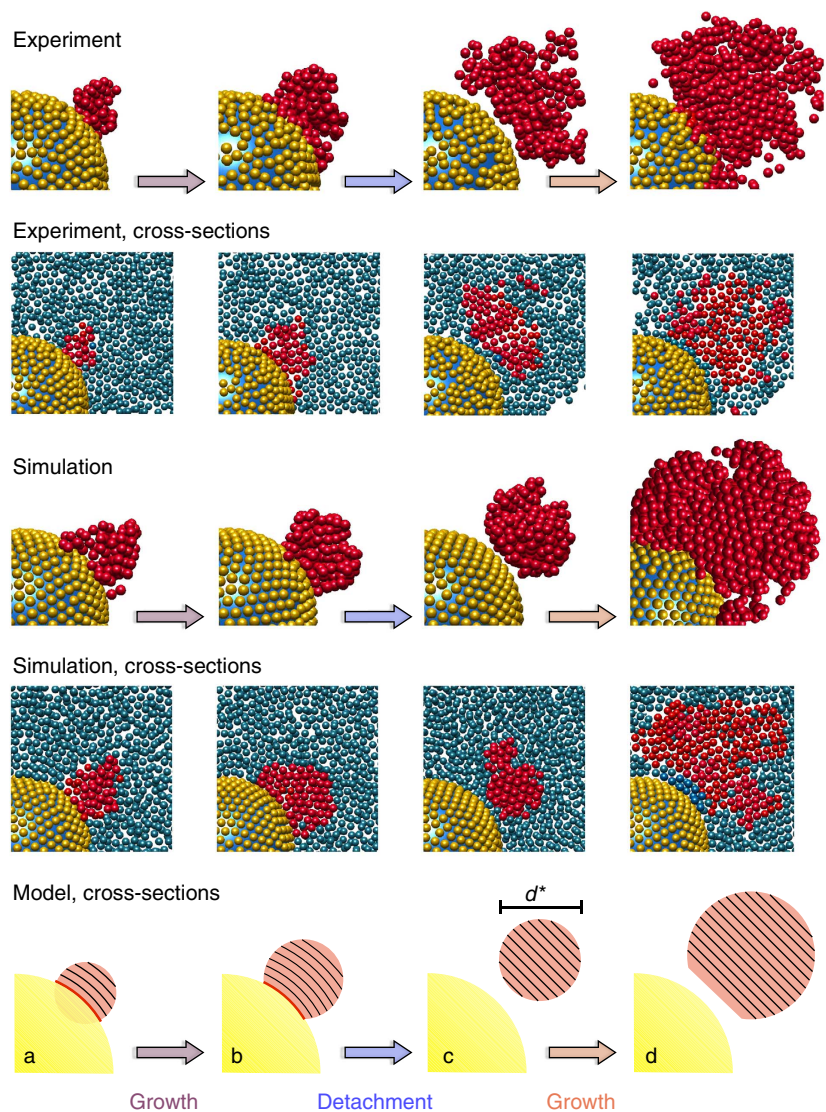

**Figure 2 | Crystallization in the presence of a seed proceeds in several stages.** (**a**) Heterogeneous crystallization on the surface of the seed, (**b**) growth of the crystallite with distorted crystal planes, (**c**) detachment of the crystallite once it reaches the critical diameter $d^*$ and (**d**) growth of the detached and relaxed crystallite in the bulk, except for a thin fluid layer between the crystallite and large sphere. Snapshots are shown as three-dimensional representations and cross-sections at times $t = 104, 260, 468, 546\tau_B$ (experiments) and $10, 40, 140, 300\tau_B$ (simulations). Crystalline particles belonging to the same cluster are represented as red spheres, whereas fluid particles are shown in blue (cross-sections) or are omitted (three-dimensional representation). Particles covering the seed are represented in yellow. In the experiments and simulations, the size ratio of large to small spheres $\sigma_2/\sigma_1 = 15$. In addition, the different stages are schematically illustrated in a cross-sectional view showing the seed (yellow) and crystallite (red) with its crystal planes (lines). Note that the colours of the arrows are used to identify the different stages in the following figures.

Whereas the time dependence of $N_c(l,t)/N$ reflects the separation into different stages (Fig. 3a,b), the individual stages are characterized by the radial, that is, layer number $l$, dependence of $N_c(l,t)/N$ (Fig. 3c–h). Initially (Fig. 3c,f), the fraction of crystalline particles increases, first at small and then mainly at larger distances, around $l = 4$–6 layers, from the surface. This is consistent with our qualitative observation (Fig. 2) that the crystallite heterogeneously grows from the seed surface. Subsequently (Fig. 3d,g), $N_c(l,t)/N$ decreases close to the seed surface, where now fluid particles dominate, whereas the fraction of crystalline particles still seems to increase further from the seed. This signals the detachment of the crystallite from the seed surface and the formation of a liquid region between the large sphere and crystallite, and, additionally, the growth of the crystallite towards the bulk. In addition, the maximum of $N_c(l,t)/N$ shifts to larger $l$, also indicating the growth of the crystal. Finally (Fig. 3e,h), $N_c(l,t)/N$ increases again. This increase begins and is most evident at a distance from the large sphere, demonstrating that the crystallite grows in the bulk. In contrast, next to the large sphere, that is, in the first about two layers, $N_c(l,t)/N$ stays very small, indicating that a thin fluid layer remains. This implies that in contrast to the beginning, now the large sphere acts as impurity that impedes crystallization in its vicinity. The analysis of $N_c(l,t)/N$ hence provides detailed information on the whole process and both the experimental and simulation results support the qualitative picture presented above (Fig. 2).

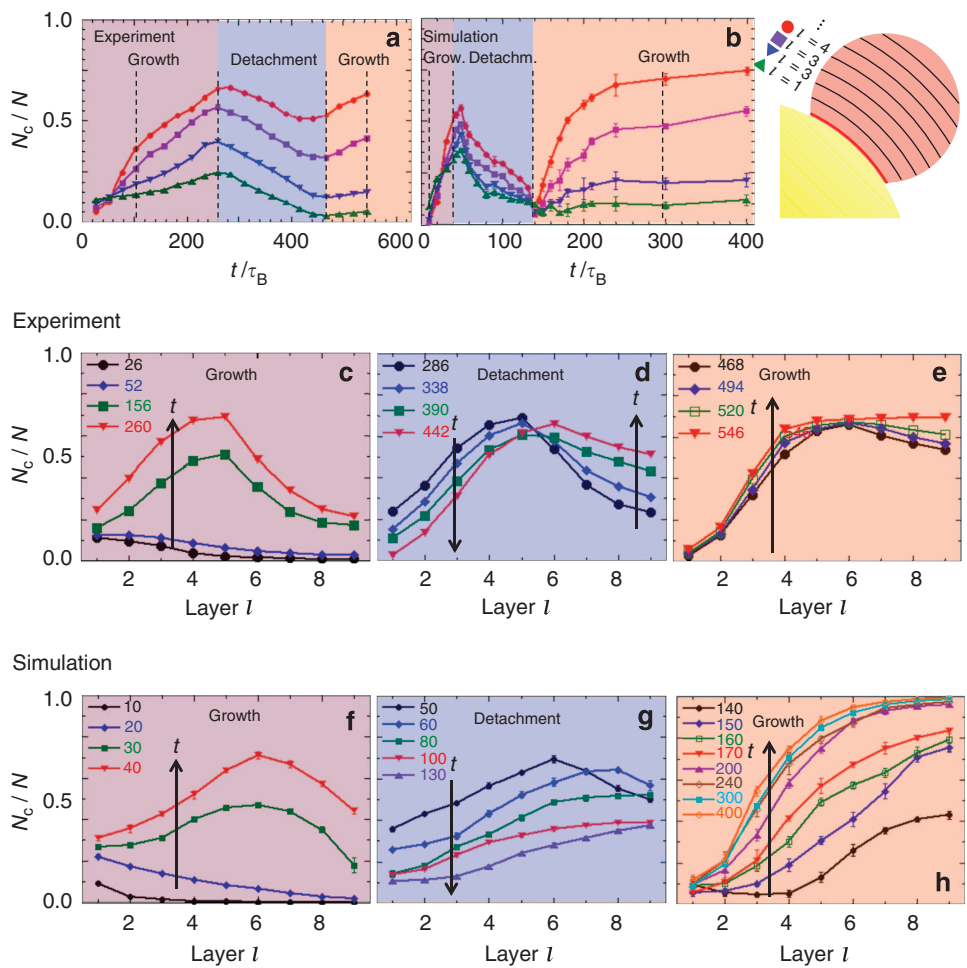

**Figure 3 | Separation and characterization of the different stages.** (**a**,**b**) Time dependence of the fraction of crystalline particles in the cluster, $N_c(l,t)/N$, in the first layers surrounding the seed (from bottom to top) as observed in experiments (**a**) and simulations (**b**). Times shown in Figs 2, 4 and 5 are indicated by vertical dashed lines. (**c**–**h**) Dependence of $N_c(l,t)/N$ on the radial distance from the seed surface, that is, layer number $l$, for the times $t/\tau_B$ indicated as observed in experiments (**c**–**e**) and simulations (**f**–**h**). The individual plots correspond to the different stages introduced in Fig. 2. Size ratio $\sigma_2/\sigma_1 = 15$.

**Particle dynamics indicate transient fluid.** A characteristic feature of the proposed process is the detachment of the crystallite. This implies the formation of a fluid between the crystallite and seed, and the subsequent refreezing of the fluid, except very close to the large sphere. The above argument for this scenario is based on structural information, in particular the fraction of crystalline, $N_c/N$, and fluid, $1 - N_c/N$, particles (Fig. 3). To complement this structural evidence, the dynamics was investigated by simulations, whereas confocal microscopy does not provide the necessary time resolution. The mean squared displacement $\langle \Delta r^2(\tau) \rangle$ after a lag time $\tau$ was determined for subsequent layers $l$ surrounding the large sphere, that is, for increasing distance from the seed surface (Fig. 4). The mean squared displacement characterizes the local particle mobility, which is much larger in a fluid than in a crystal.

Initially ($t = 10\tau_B$), the mean squared displacement $\langle \Delta r^2(\tau) \rangle$ increases with increasing distance $l$ from the large sphere. Far from the seed (layers $l = 7$ and 8, red circles in Fig. 4a), the diffusion coefficient reaches a value, $D_f \approx 1.5 \times 10^{-3} D_0$, as expected[37] for a fluid with the present volume fraction, that is, $\Phi = 0.53$, and decreases towards the surface due to the surface-induced order. This suggests the presence of fast diffusing, that is, fluid, particles further from the seed and a larger fraction of crystalline, and thus less mobile particles close to the seed. At the time when the maximum in $N_c(l,t)/N$ of the first layers is observed (Fig. 3b), that is, just before the crystallite detaches ($t = 40\tau_B$), indeed small $\langle \Delta r^2(\tau) \rangle$ are observed at all distances $l$. After detachment ($t = 140\tau_B$), $\langle \Delta r^2(\tau) \rangle$ increases towards the large sphere. Close to the large sphere, the diffusion coefficient is similar to the initial diffusion coefficient $D_f$ of the fluid particles. It is noticeable that in this case the proximity of the seed surface reduces the mobility slightly[38,39], namely of the first two layers ($l = 1$ and 2, green triangles in Fig. 4c). This is consistent with the existence of a fluid between the crystallite and large sphere. Subsequently ($t = 300\tau_B$), the dynamics slows down with a significant decrease in $\langle \Delta r^2(\tau) \rangle$ further from the large sphere, whereas the particles close to the seed remain mobile, again with a diffusion coefficient similar to the diffusion coefficient $D_f$ of the fluid particles. This indicates the freezing of the fluid, except for an about two-layer-thin fluid next to the large sphere.

In addition to the structural findings, the evolution of the dynamics (Fig. 4) hence provides further, and notably, independent support for the proposed mechanism of crystal growth, detachment and melting, and finally refreezing of the transient fluid, except for a thin fluid layer next to the large sphere. It is noteworthy that during the whole process ($\tau \lesssim 300\tau_B$), the displacements of the individual particles are very small with $\langle \Delta r^2(300\tau_B)\rangle \ll \sigma_1^2$. The rearrangements, including the detachment of the crystallite, are thus based on structural changes while the actual movements of the individual particles remain very limited.

**Strength of the distortions.** Having presented structural and dynamic evidence for the detachment of the crystallite, we investigate the driving force for the detachment: accumulated elastic stress due to distortions of the crystal planes. We characterize the distortions by the averaged local bond-orientational order parameters $\bar{q}_4$ and $\bar{q}_6$ of the crystalline particles[40]. The $\bar{q}_4$ and $\bar{q}_6$ values of the particles forming the crystallite are determined in the experiments and simulations (Fig. 5). They can be compared with the values of the limiting cases, namely an fcc crystal (upright triangle) and a liquid (inverted triangle).

Initially, the $\bar{q}_4$ and $\bar{q}_6$ values indicate a distorted crystallite (a,e,i), which at this time is still small with only about two layers of crystalline particles surrounding the seed (green and blue). The subsequently crystallized and hence more distant layers (violet to yellow) have a smaller curvature and are indeed less distorted (b,f,j). Moreover, they impose additional distortions on the subjacent layers, indicated by their $\bar{q}_4$ and $\bar{q}_6$ values moving towards smaller values. After the detachment, the crystal is relaxed and an fcc structure prevails (c,g,k). This crystallite involves significantly less particles, which are mainly located further from the seed (towards yellow). The crystal subsequently grows (d,h,l). Although growth also proceeds towards the large sphere (orange and red), it hardly involves particles next to the large sphere (green and blue). In addition, closer to the large sphere small distortions are noticeable, whereas the bulk of the crystal (yellow) shows an fcc structure.

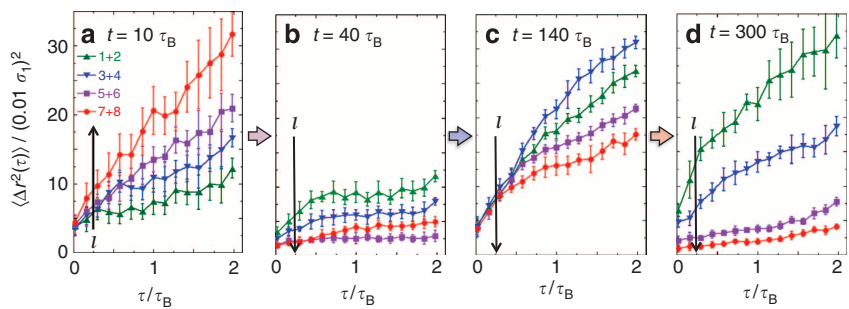

**Figure 4 | Evolution of the particle mobility.** The mean squared displacement $\langle\Delta r^2(\tau)\rangle$ as a function of delay time $\tau$ for different times (**a**) $t=10$, (**b**) 40, (**c**) 140 and (**d**) $300\tau_B$ after the start of crystallization. It reveals an initial reduction in mobility, indicating freezing (a→b), then enhanced diffusion pointing at melting (b→c) and again a reduced mobility, suggesting the refreezing of the transient fluid, except for a thin fluid layer next to the large sphere (c→d). The legend indicates layer numbers $l$, that is, distances from the seed surface. The arrows point towards larger distances $l$. Size ratio $\sigma_2/\sigma_1 = 15$. The data were obtained by simulations.

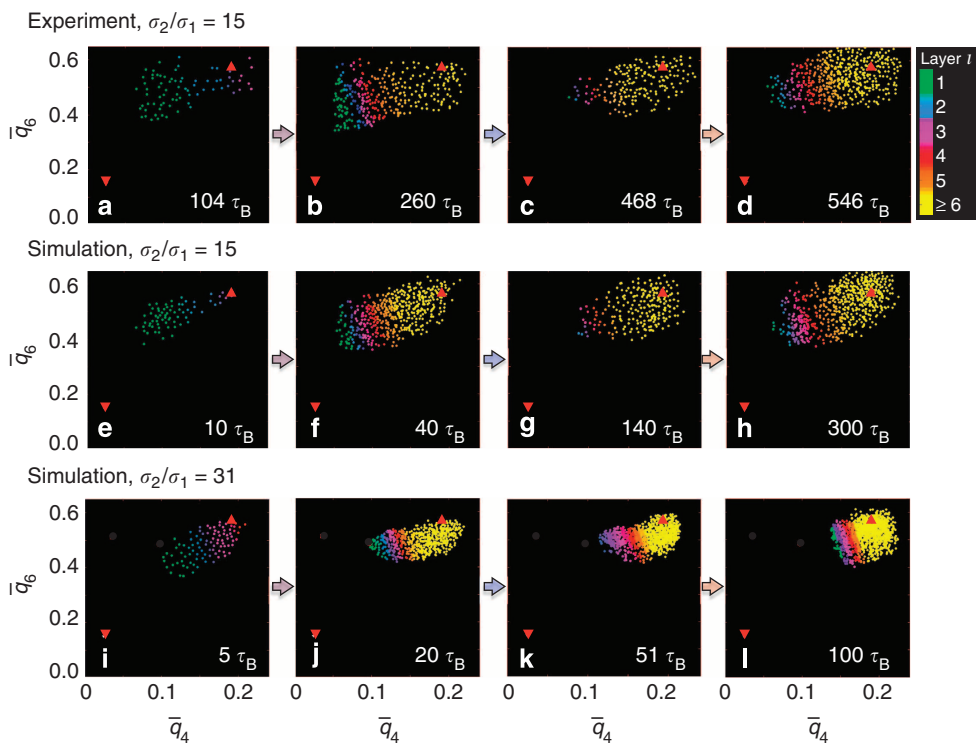

**Figure 5 | Curved seeds lead to distortions.** The crystalline particles' averaged local bond-orientational order parameters $\bar{q}_4$ and $\bar{q}_6$ characterize local distortions. For the two limiting cases, namely an ordered fcc crystal and a disordered fluid at temperature $T=0$, the $\bar{q}_4$ and $\bar{q}_6$ values are indicated by upright and inverted red triangles, respectively. The $\bar{q}_4$ and $\bar{q}_6$ values of individual particles are determined by (**a**–**d**) experiments and (**e**–**l**) simulations, and represented by points whose colours indicate the layer number $l$. The times $t$ after the start of crystallization are given and correspond to the characteristic times illustrated in Fig. 3a,b, that is, (**a,e,i**) short time, (**b,f,j**) maximum in $N_c(l,t)/N$, (**c,g,k**) minimum in $N_c(l,t)/N$ and (**d,h,l**) long time. They show that the crystal planes are distorted before the detachment (**a,b;e,f;i,j**), relax after the detachment (**c,g,k**) and subsequently the relaxed crystal continues to grow (**d,h,l**). Size ratios $\sigma_2/\sigma_1 = 15$ (**a–h**) and 31 (**i–l**).

With increasing diameter $\sigma_2$ of the large sphere, and thus a smaller curvature of its surface, the distortions become less pronounced, as indicated by $\bar{q}_4$ and $\bar{q}_6$ values that are closer to the values of an undistorted fcc crystal for a given layer number $l$ (Fig. 5i–l). Still, for seed sizes $\sigma_2 \leq 31\sigma_1$, the $\bar{q}_4$ and $\bar{q}_6$ values characteristic for an undistorted fcc crystal are only observed in the outermost layers $l \geq 6$ (yellow).

The experimental and simulation results both suggest that the crystal planes of a crystallite, which is heterogeneously grown on a curved seed, are distorted, especially close to the seed. The distortions appear to drive the detachment, which allows the crystal to relax. This is consistent with the observation that no detachment occurs for very large and hence less-curved seeds.

**Theoretical description of detachment.** The diameter $d(t)$ of the growing crystallite, quantified by twice its radius of gyration, is followed in the experiments and simulations for different seed sizes $\sigma_2$. Owing to the large sizes of the crystallites, $d^3 \gg \sigma_1^3$, typically only single crystallites can be observed in individual experiments or simulations, which are, however, repeated to accumulate statistics. Complete detachment is indicated by the minimum in the fraction of crystalline particles $N_c(l,t)/N$ close to the seed, that is, small $l$ (Fig. 3a,b). The diameter of the crystallite at this time is taken as its critical diameter $d^*$ (Fig. 2c). The critical diameter $d^*$ is found to increase with the seed size $\sigma_2$ (Fig. 6). The experimental and simulation results agree within their statistical uncertainties. The slightly larger $d^*$ observed in experiments are attributed to possible differences in the volume fractions[36], interaction potentials[33] and polydispersities. The observed increase of the critical size $d^*$ with increasing $\sigma_2$, that is, decreasing seed curvature, can be related to the decreasing distortions (Fig. 5e–h versus Fig. 5i–l) and thus a decreasing driving force for detachment, which results in the prolonged accumulation of (smaller) elastic stresses before detachment occurs. In addition, for the investigated conditions, namely volume fraction $\Phi = 0.53$, particle–particle interactions and polydispersity, no detachment is observed for very large seeds, $\sigma_2 > 31\sigma_1$, and thus very small distortions and elastic stresses (Fig. 6, hatched area).

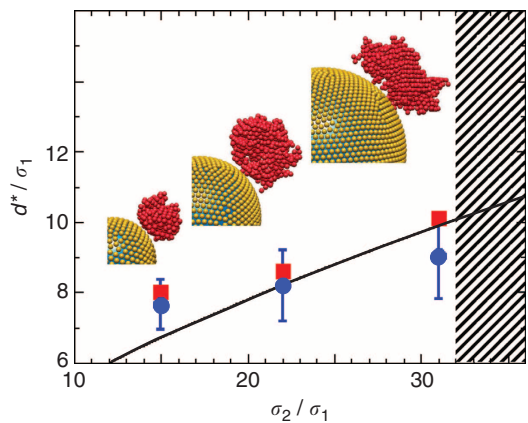

**Figure 6 | Distortions lead to the detachment of the crystallite.** Critical diameter $d^*$ of the detaching crystallite as a function of the seed diameter $\sigma_2$ normalized to the particle diameter $\sigma_1$ as observed in experiments (red squares) and simulations (blue dots). Simulation snapshots show the corresponding crystallites. The line represents calculations (equation (3)) without any free parameters. The hatched area on the right indicates seed sizes for which no detachment was observed.

To quantify the relation between $d^*$ and $\sigma_2$, we develop a theoretical model for the detachment of the crystallite. The detachment is controlled by a competition between the elastic energy penalty, $F_e$, of the distorted crystallite[41,42] and the energy, $F_i$, required to create additional interfaces during detachment. Detachment occurs when $F_e = F_i$. We want to focus on this balance, which we expect to represent the fundamental underlying physics, and hence introduce a few simplifications. Based on our experimental and simulation observations (Fig. 2), it appears reasonable to approximate the shape of the crystallite by a spherical cap, which is also suggested by the classical theory of heterogeneous nucleation[43]. The cap is found to be small compared with the seed, that is, $d \leq d^* \ll \sigma_2$ (Fig. 2), and to detach without significant volume change, that is, without a sudden change in its number of particles (Fig. 3a,b). Furthermore, the observed $\bar{q}_4$ and $\bar{q}_6$ values suggest that at detachment the layers most distant from the seed ($l \geq 6$) can be considered to have only small distortions and almost an fcc structure (Fig. 5b,f,j).

Under these conditions, elasticity theory predicts for the elastic stress[44]

$$F_e = \frac{E}{24} \frac{d^5}{(\sigma_2 + 2d)^2} \qquad (1)$$

with $E$ being the bulk modulus of the crystal. For an fcc crystal and a volume fraction $\Phi = 0.53$, a reasonable estimate is $E = 30 k_B T/\sigma_1^3$ with $k_B T$ being the thermal energy[45]. The interface energy $F_i$ takes into account the creation of new fluid–crystal and fluid–seed interfaces to replace the seed–crystal interface:

$$F_i = \frac{\pi}{4} d^2 \Delta\gamma. \qquad (2)$$

The difference in interface tensions has been shown[46–48] to be about $\Delta\gamma = 0.6 k_B T/\sigma_1^2$.

The crystallite detaches once the elastic energy penalty $F_e$ reaches the interfacial energy gain, $F_i$, that is, $F_e = F_i$. This leads to a prediction for the critical diameter $d^*$ in dependence of the seed radius $\sigma_2$,

$$\sqrt{E'} \left(\frac{d^*}{\sigma_1}\right)^{3/2} - 2\left(\frac{d^*}{\sigma_1}\right) = \left(\frac{\sigma_2}{\sigma_1}\right), \qquad (3)$$

where $E' = E\sigma_1/(6\pi\Delta\gamma) = 2.7$ with the above values. Hence, this prediction does not contain any free parameters. Despite this, the model predicts the observed magnitude of $d^*$ correctly and reproduces the dependence on the seed size $\sigma_2$ about right (Fig. 6). This supports our suggestion that the crystallite accumulates elastic stress due to seed-induced distortions and detaches once the accumulated elastic stress balances the interfacial energy gained through the heterogeneous nucleation on the seed.

Equation (3) suggests that the crystallite continues to grow to larger sizes $d^*$ for larger seed sizes $\sigma_2$ or, equivalently, smaller seed curvatures and hence smaller seed-induced distortions. Only for flat walls, that is, $\sigma_2 \to \infty$, and thus in the absence of any seed-induced distortions, the crystallite is predicted to remain on the seed. The crystallite might, however, also not detach if the elastic stress can be relaxed through other mechanisms, such as defects, polydispersity, anisotropic particle shape or softness[28]. Whereas these effects are beyond the scope of the present work, a kinetic effect is considered. The crystallite must transform from a distorted to a relaxed and detached state within a reasonable time. This transition occurs through fluctuations, whose typical energy is

$$F_f = 2\sqrt{N} k_B T \quad (4)$$

where $N = \Phi(d/\sigma_1)^3$ is the number of particles forming the crystallite and the prefactor 2 is based on an ideal gas estimate. For detachment to occur, this energy must be larger than the interfacial or elastic energy, which are identical at detachment; $F_f > F_e$ ($= F_i$ at detachment). This results in a maximum diameter $d^*_{max}$ of the crystallite, beyond which it will not detach:

$$\frac{d^*_{max}}{\sigma_1} = \frac{64\Phi}{\pi^2}\left(\frac{k_B T}{\Delta\gamma\sigma_1^2}\right)^2. \quad (5)$$

Owing to the relation between $d^*$ and $\sigma_2$ (equation (3)), this implies that detachment becomes kinetically hindered beyond a maximum seed diameter:

$$\frac{\sigma_{2,max}}{\sigma_1} = \sqrt{E'}\frac{512\Phi^{3/2}}{\pi^3}\left(\frac{k_B T}{\Delta\gamma\sigma_1^2}\right)^3 - \frac{128\,\Phi}{\pi^2}\left(\frac{k_B T}{\Delta\gamma\sigma_1^2}\right)^2. \quad (6)$$

For the present system, the values given above yield $\sigma_{2,max}/\sigma_1 \approx 29$. This is consistent with our observation that no detachment occurs for seeds with $\sigma_2/\sigma_1 > 31$. The consistency of this prediction with the experimental and simulation findings lends further support to the proposed mechanism.

## Discussion

We have quantitatively and on the single-particle level investigated crystallization in the presence of a seed by combining confocal microscopy, computer simulations and theory. A multi-step scenario was identified (Fig. 2, bottom). Crystallization is initiated by the seed. However, owing to a mismatch between the thermodynamically favoured and seed-induced crystal structure, here curved lattice planes, distortions and thus increasing elastic stresses accumulate. Once the elastic energy penalty reaches the interfacial energy gain, it is energetically favourable for the crystallite to detach from the seed and release the elastic stress, although this requires to create additional interfaces. For this process to occur, the energy of the fluctuations must exceed the elastic or interfacial energy, which are identical at detachment. The detached and relaxed crystallite then continues to grow, but now in bulk. Thus, heterogeneous nucleation and growth is followed by crystal growth in bulk. Furthermore, this implies that between the seed and crystallite, the crystal melts and a fluid first forms and then refreezes, except a thin fluid layer that remains next to the large sphere. The 'seed' no longer favours crystallization but now acts as impurity.

To be able to quantitatively tune the mismatch and to visualize and follow crystallization on the single-particle level, we used a colloidal model system. Nevertheless, any system in which the heterogeneously grown crystal has a different structure from the thermodynamically favoured crystal will accumulate seed-induced distortions. They not only result in a decrease of the crystallization speed[8,22,28–30] but can also lead to detachment allowing the crystallite to release the elastic stress. This is expected to be very common, as a perfect match is very difficult to achieve in practice. Furthermore, the thermodynamically favoured crystal structure might even be unknown or, for material property reasons, not desired. The scenario presented here, therefore, is of relevance for many industrial processes as well as the rational design of crystalline or partially crystalline materials.

The materials might be, as in the present study, colloidal, for example, for photonics applications[13], or consist of metal atoms[10] or small molecules[11], which have been found to follow very similar principles[49]. The details of the detachment process, however, may depend on the actual system considered. For example, materials containing soft particles may be able to partially or completely release accumulated distortions through their softness, binary or polydisperse systems through variations of the local composition, and anisotropic particles, such as in liquid crystals, through their additional rotational degrees of freedom. On the other hand, in polydisperse and anisotropic systems, the seed-induced crystallite might have a different local composition or orientation, respectively, than the thermodynamically favoured bulk crystal, which might contribute an additional mechanism to accumulate stress. Similar arguments apply to atomic and molecular systems, such as metals or protein solutions, in which, for example, the solvent may play a different role or is even absent. These effects may alter, accelerate, delay or suppress the detachment process, but in all these different systems the general scenario is expected to be similar to the one presented here.

Finally, an external field, such as gravity or flow, might draw the detached crystallite further from the seed. Then the seed is again available for heterogeneous nucleation and hence might repeatedly initiate heterogeneous nucleation; it acts as crystallization catalyst[22]. To explore these aspects and to unravel the details of the material-specific processes underlying heterogeneous crystallization, further concerted efforts combining real-space experiments, computer simulations and analytical theories are needed.

## Methods

**Experiments.** Fluorescently labelled and sterically stabilized PMMA particles with diameter $\sigma_1 = 1.83\,\mu m$ (as determined by static light scattering) and a polydispersity of $<4\%$ (as determined by dynamic light scattering) are dispersed in a mixture of cis-decalin and cycloheptylbromide, which matches the density and approximates the refractive index of the particles. In this solvent mixture, the particles acquire a small charge. To screen this charge, the solvent mixture was saturated with salt (tetrabutylammoniumchloride)[33,50,51], which results in a Debye screening length below 100 nm[33,50,51]. The particles hence exhibit hard-sphere-like behaviour[33]. Thus, the effective volume fraction might be slightly larger than the volume fraction $\Phi = 0.53$ (refs 33,36). For this system, the Stokes–Einstein relation predicts a Brownian time $\tau_B = \sigma_1^2/D_0 = 33\,s$ with the short-time infinite-dilution diffusion coefficient $D_0$. A small number of large spherical glass beads are added ($28\,\mu m \leq \sigma_2 \leq 400\,\mu m$; supplied by Whitehouse Scientific) whose diameter $\sigma_2$ was determined by confocal microscopy. Owing to their large density, they settle to the bottom of the container. During the measurements, they move slightly with respect to the observation volume, typically far below $1\,\mu m$, which is taken into account in the data analysis.

The samples are kept and imaged in glass vials whose bottoms are replaced by microscope cover slips[52], which are coated with polydisperse PMMA particles (diameters between 3 and $8\,\mu m$), to prevent heterogeneous nucleation. These vials allow stirring the samples before each measurement, with a small magnetic bar that, to avoid damage to the coating layer, is kept floating by a home-built device. Despite intense mixing, between one and five layers next to the glass beads remain partially ordered, which is attributed to favourable wetting[7,53].

The confocal head (Olympus FluoView FV1000) was attached to an inverted microscope (Olympus IX81) with a $\times 60$ oil-immersion objective (numerical aperture 1.35). Starting from the cover slip, observation volumes of $212 \times 212 \times 50\,\mu m^3$ are scanned, which correspond to 251 slices with a distance of 200 nm and a size of $1,024 \times 1,024$ pixels per slice. An observation volume contains about 350,000 particles. A single slice requires about 3 s and a stack of slices requires about 15 min to scan. The whole observation volume was scanned every 15 min, that is, about every $26\tau_B$, for about 5 h, which allows us to follow crystallization, including detachment. However, a sufficiently large volume cannot be scanned fast enough to determine the mean squared displacement $\langle\Delta r^2(\tau)\rangle$ with the required time resolution, that is, $\Delta\tau = \tau_B/7 < 5\,s$ (Fig. 4).

Only particles at least $2\,\mu m$ from the cover slip are considered in the data analysis, to avoid wedge effects between the cover slip and the glass beads. The particle coordinates are determined using standard procedures[34] and are estimated to have an accuracy of about 30–100 nm depending on the local image quality, mainly depth in the sample[33,52]. The particles are assigned to layers of width $0.91\sigma_1$, which corresponds to the separation of the crystal planes, around the large spheres, taking into account the above mentioned slight movement of the large spheres.

**Simulations.** Brownian dynamics simulations are performed that use a spherical simulation box of diameter $\sigma_0$. It is filled with 500,000 hard spheres with diameter $\sigma_1$ and volume fraction $\Phi = 0.53$, as well as a large hard sphere with diameter $\sigma_2$ fixed in the centre of the box. To mimic the precrystallized layer in the experiments, the large sphere is decorated by a hexagonal layer of fixed hard spheres with a lattice constant $1.133\sigma_1$ and, unavoidably, a few defects. If the large sphere is decorated by a layer of particles on which no order is imposed or which only initially are hexagonally arranged but then allowed to move freely, qualitatively similar but quantitatively different behaviour is observed. The outer wall of the simulation box is covered with fixed disordered spheres, which interact with the mobile spheres via a Yukawa potential whose range is adjusted to yield a constant volume fraction of the fluid during the whole crystallization process. This is similar in spirit to a constant pressure simulation and mimics the experimental situation.

An initial configuration of non-overlapping randomly arranged particles is prepared by random insertions of spheres. The initial configuration does not significantly affect the results apart from rearrangements during a time $t < \tau_B$. Simulations were run with time steps of about $0.001\tau_B$. To improve statistics, several runs, typically three, are performed with different initial configurations. The results are averaged and the s.d. determined and shown as error bars. In addition, to improve the statistics of the mean squared displacement $\langle \Delta r^2(\tau) \rangle$ (Fig. 4), runs with three different initial times $t - \Delta t$, $t$ and $t + \Delta t$, with $\Delta t = \tau_B \ll t$, are averaged.

**Data analysis.** In experiments and simulations, crystalline particles are identified based on the local bond-orientational order parameter $q_6$[35], where

$$q_{lm}(i) = \frac{1}{N_b(i)} \sum_{j=1}^{N_b(i)} Y_{lm}(\mathbf{r}_{ij}) \tag{7}$$

$$q_l(i) = \left( \frac{4\pi}{2l+1} \sum_{m=-l}^{l} |q_{lm}(i)|^2 \right)^{1/2} \tag{8}$$

$$\mathbf{q}_l(i) \cdot \mathbf{q}_l(j) = \sum_{m=-l}^{l} q_{lm}(i) q_{lm}(j)^* \tag{9}$$

with $N_b(i)$ being the number of nearest neighbours of particle $i$, $Y_{lm}$ the spherical harmonics and $\mathbf{r}_{ij}$ a unit vector in the direction of the bond between particle $i$ and its neighbour $j$.

Particles are declared to be neighbours if their centres are within $1.17\sigma_1$. Two neighbouring particles $i$ and $j$ are considered connected in a crystallite if their orientational order parameters fulfill $\mathbf{q}_6(i) \cdot \mathbf{q}_6(j) > 0.5$. If a particle has at least eight connected neighbours, it is regarded a crystalline particle. Finally, crystalline particle clusters consist of crystalline particles that are at most $1.03\sigma_1$ away from the particles with which they share a Voronoi surface. This is implemented similar to the Stoddard algorithm[54].

The centre and maximum extent of the cluster at the time when $N_c(l,t)/N$ reaches its first maximum (Fig. 3a,b) is used to define a spherical region with the same centre and maximum extent. Within this region, the time and radial dependence of the fraction of crystalline particles $N_c(l,t)/N$ (Fig. 3), mean squared displacement $\langle \Delta r^2(\tau) \rangle$ (Fig. 4), as well as $\bar{q}_4$ and $\bar{q}_6$ values (Fig. 5) are calculated. Their radial dependences are represented in layers with a width of $0.91\sigma_1$, which corresponds to the separation of the lattice planes.

### Acknowledgements
We thank A.B. Schofield (The University of Edinburgh) for supplying PMMA particles, the Centre for Advanced Imaging (University Düsseldorf) for time on the confocal microscope and the DFG for financial support (SPP 1296).

### Author contributions
S.U.E. and H.L. designed the research. E.A. performed the simulations. K.S. conducted the experiments. All authors contributed to the analysis and interpretation of the results, and to the writing of the paper.

### Additional information
**Competing financial interests:** The authors declare no competing financial interests.

**Reprints and permission** information is available online at http://npg.nature.com/reprintsandpermissions/

**How to cite this article:** Allahyarov, E. *et al.* Crystallization seeds favour crystallization only during initial growth. *Nat. Commun.* 6:7110 doi: 10.1038/ncomms8110 (2015).